\documentclass[aip,cha, preprint,showpacs,showkeys]{revtex4-1}
\usepackage{amssymb, dcolumn,  graphicx, latexsym, amsfonts, bm}

\begin{document}
\title{The ion acoustic instability of the cylindrical inhomogeneous helicon 
discharge plasma with rotating electrons}
\author{V. V. Mikhailenko}\email[E-mail:]{vladimir@pusan.ac.kr}
\affiliation{BK21 FOUR Information Technology, Pusan National University,  Busan 46241, 
South Korea}
\author{Hae June Lee}\email[E-mail:]{haejune@pusan.ac.kr}
\affiliation{Department of Electrical Engineering, Pusan National University, 
Busan 46241, South Korea}
\author{V. S. Mikhailenko}
\affiliation{Plasma Research Center, Pusan National University,  Busan 46241, South Korea}
\author{M. O. Azarenkov}
\affiliation{V.~N. Karazin Kharkiv National University, Kharkiv 61022, Ukraine}
\date{\today}

\begin{abstract} 
The kinetic theory of the microinstabilities of a cylindrical plasma, produced by the 
cylindrical azimuthally symmetric (azimuthal mode number $m_{0}=0$) helicon wave, is 
developed. This theory is based on the derived linear integral equation for the 
Fourier-Bessel transform 
of the electrostatic potential, which accounts for the plasma response on the macroscale 
radial inhomogeneity of the helicon wave, which is commensurable with radial 
scale of the plasma density inhomogeneity, and on the microscale, which is 
commensurable with the thermal Larmor radius of electrons. The developed theory reveals 
new macroscale effect of the azimuthal steady rotation of electrons with a radially 
inhomogeneous angular velocity, caused by the 
radial inhomogeneity of the helicon wave. The solution of the integral equation for 
the electrostatic potential, derived in the short-wavelength limit, is derived 
in the form of the the functional equation for the electrostatic potential, 
coupled with infinite number of its satellites at a frequency separation equal 
to the frequency of the helicon wave. It is the basic 
equation for the investigations of the dispersion properties of the parametric and 
current driven instabilities of the cylindrical plasma in the radially inhomogeneous helicon wave. 
The analytical solution of the derived dispersion  equation is found for the high frequency 
kinetic ion acoustic instability of the cylindrical helicon plasma, driven by the coupled effect 
of the electron diamagnetic drift and of the steady azimuthal rotation of electrons relative 
to the ions with a radially inhomogeneous angular velocity.
\end{abstract}

\maketitle

\section{Introduction}\label{sec1}
The helicon plasma sources attract great interest in plasma community and have various 
applications\cite{Shinohara, Takahashi} due to the remarkably strong absorption of helicon waves 
in plasmas and anomalously strong electron heating\cite{Boswell}. The helicon discharge makes use 
of the helicon wave which is inductively launched into the plasma column.
The linear theory of the helicon wave, which is the whistler wave in a bounded plasma, predicts that 
this wave has phase velocity much above the electron thermal velocity and, therefore, the absorption 
of the helicon wave in the collisionless plasma by electrons due to the 
electron Landau damping is a negligibly weak. That is why the experimentally observed
\cite{Boswell} unusually high absorption rate of the helicon wave, that testified to strong 
interaction of the helicon wave with electrons, was unpredictable by the linear theory 
of the helicon wave propagation. 
Although a large number of studies have been carried out, the mystery of why helicon discharges 
are so efficient is still unresolved. 

The anomalous absorption of helicons and plasma heating was observed a very long time ago in 
the first experimental studies\cite{Grigor'eva, Porkolab} of the basic plasma physics processes. 
It was claimed in these papers that the anomalous absorption of a large amplitude whistler wave
was caused by the development of the current driven ion acoustic instability\cite{Grigor'eva} or by 
the development of the resonant decay instability\cite{Porkolab}. The Boswell's experiments 
gave impetus to the active theoretical investigations of the plasma instabilities driven by the 
helicon wave. It was found that the development of the 
parametric kinetic\cite{Akhiezer, Mikhailenko4} and decay\cite{Aliev,Lorenz,Kramer} ion acoustic instabilities, 
originated from the oscillatory motion of electrons relative to 
ions in the pumping helicon field, may be the cause of the anomalous absorption of the helicon wave 
and of the anomalous heating of electrons, resulted from the interaction of electrons with 
ion acoustic turbulence. Since then, the plasma turbulence in helicon plasma 
was investigated in several experiments \cite{Lorenz,Altukhov,Kramer} in which the 
detected short scale fluctuations were identified as the ion acoustic\cite{Kaganskaya} and 
Trivelpiece - Gould\cite{Lorenz} waves, and it was found that the level of these fluctuations 
increases with RF power. 

The theory of the parametric instabilities of the helicon sources plasma is developed 
to date for a model of a slab plasma\cite{Akhiezer, Aliev, Kramer} in the field of 
spatially uniform electric field $\mathbf{E}\left(t\right)$ of the 
helicon pump wave. However, the helicon wave field in cylindrical 
helicon sources is as a rule spatially inhomogeneous with radial inhomogeneity length 
comparable with, or less than, the radius of plasma cylinder. The effect of the cylindrical 
geometry of the plasma and of the helicon wave field, and the effect of the spatial 
inhomogeneity of the helicon wave on the parametric microturbulence is usually ignored assuming 
that the model of the uniform electric field oscillated with the helicon wave frequency 
is sufficient for the proper description 
of the parametric instabilities the wavelengths of which are much less than the radial 
inhomogeneity scale length of the helicon wave in plasma cylinder. It was found in Refs.
\cite{Mikhailenko, Mikhailenko1}, however, that the electromagnetic field inhomogeneity 
in the inductive plasma sources may be the powerful source of the instabilities development. 
It was derived\cite{Mikhailenko, Mikhailenko1} 
that the accelerated motion of electrons relative to ions under the action of the 
ponderomotive force, formed in the skin layer of the inductively coupled plasma, 
may be the much more stronger source of the 
instabilities development than the quiver motion of the electron in the electromagnetic field.
The effects of the spatial inhomogeneity of the helicon wave field 
on the anomalous absorption of a helicon wave in the real helicon sources was not investigated yet.
The spatial structure of the helicon wave in the helicon sources depends on the 
antenna design and on the distance from the antenna, on the input RF power and RF frequency, 
on the magnitude and radial profile of the electron density, etc.\cite{Chen, Chen1, Aliev1}. 
The focus of this paper is the development of the kinetic theory of the 
microscale instabilities of the radially inhomogeneous cylindrical plasma, driven by the 
cylindrical azimuthally symmetric $(m=0)$ radially inhomogeneous 
helicon wave. In Sec. \ref{sec2}, we present the basic equations of our Vlasov-Poisson 
theory of the stability of the cylindrical plasma in the field of the azimuthally 
symmetric radially inhomogeneous helicon wave. 

The main result of Sec. \ref{sec3} is the derived integral 
equation for the separate azimuthal mode of the Fourier-Bessel transform of the electrostatic 
potential. The solution to this equation is found for the microscale short-wavelength 
perturbations in the form of the functional equation for the electrostatic potential, 
coupled with infinite number of its satellites with a frequency separation equal 
to the harmonics of the helicon wave frequency. This equation governs the 
dispersion properties of the instabilities of the radially inhomogeneous 
cylindrical plasma driven by the azimuthally symmetric inhomogeneous helicon wave.
  
The developed theory is based on the two-scale approach to the analytical 
solution of the Vlasov equation in which the microscale and macroscale responds of 
the cylindrical radially inhomogeneous plasma on 
the azimuthally symmetric radially inhomogeneous helicon wave were accounted for. 
This approach reveals new macroscale effect of the azimuthal steady 
rotation of electrons with radially inhomogeneous angular velocity, caused by the radial 
inhomogeneity of the cylindrical helicon wave. This effect is absent in the model of the 
slab plasma in the spatially uniform helicon wave. The effect of the plasma rotation was 
observed experimentally in the helicon plasma source, that used azimuthally symmetric antenna, 
a long time ago\cite{Tynan}, but was not explained yet. The theory of the short scale 
high frequency kinetic ion acoustic instability, driven by the coupled effect of the electron 
diamagnetic drift and of the rotation of electrons relative to ions, is considered 
in Sec. \ref{sec4} as a particular case of the derived functional equation. 
The Conclusions are given in Sec. \ref{sec5}.

\section{Basic equations}\label{sec2} 

We consider an axially symmetric radially inhomogeneous plasma in a uniform axial 
magnetic field $\mathbf{B}_{0}$, directed along $z$ axes, and in the electric 
$\mathbf{E}_{1}\left(r, z, t \right)$ and magnetic $\mathbf{B}_{1}\left(r,z,t \right)$ fields  
of the azimuthally symmetric helicon wave (azimuthal mode number $m_{0}=0$), excited by the loop 
antenna located on the boundary $r=r_{0}$ of the cylindrical chamber. The particle density 
near the plasma boundary is assumed to be sufficiently small for the various effects connected 
with particle collisions with the chamber walls may be neglected. In this paper, 
we consider the effect of the electron motion in the helicon wave relative 
to ions on the development of the short scale electrostatic perturbations 
with a wavelength much less than the radial inhomogeneity scale lengths of the 
plasma density, electron temperature, and of the helicon wave field. Our theory is 
based on the Vlasov equation for the electron 
distribution function $F_{e}\left(\mathbf{v}, \mathbf{r}, t\right)$, which in 
the usual cylindrical coordinates  $r, \varphi, z$
and with electron velocity components $v_{r}, v_{\varphi}, v_{z}$  directed along 
these coordinates has a form
\begin{eqnarray}
&\displaystyle
\frac{\partial}{\partial t}F_{e}\left(v_{r}, v_{\varphi}, v_{z}, r, \varphi, z, t\right)+v_{r}
\frac{\partial F_{e}}{\partial r}
+\frac{v_{\varphi}}{r}\frac{\partial F_{e}}{\partial \varphi}+v_{z}\frac{\partial F_{e}}{\partial z}
\nonumber
\\  
&\displaystyle
+\left[\frac{v^{2}_{\varphi}}{r}+\frac{e}{m_{e}}\left(E_{1r}+\tilde{E}_{r}+\frac{1}{c}
\left(v_{\varphi}B_{1z}-v_{z}
B_{1\varphi}\right)\right)
+\omega_{ce}v_{\varphi}\right]\frac{\partial F_{e}}{\partial v_{r}}
\nonumber
\\  
&\displaystyle
-\left[\frac{v_{\varphi}v_{r}}{r}-\frac{e}{m_{e}}\left(E_{1\varphi}+\tilde{E}_{\varphi}
+\frac{1}{c}\left(v_{z}B_{1r}-v_{r}B_{1z}\right)\right)
+\omega_{ce}v_{r}\right]\frac{\partial F_{e}}{\partial v_{\varphi}}
\nonumber
\\  
&\displaystyle
+\frac{e}{m_{e}}\left(\tilde{E}_{z}+\frac{1}{c}\left(v_{r}B_{1\varphi}-v_{\varphi}B_{1r}\right)\right)
\frac{\partial F_{e}}{\partial v_{z}}=0,
\label{1}
\end{eqnarray}
where $e<0$ is the electron charge,  $\omega_{ce}=eB_{0}/m_{e}c$ is the electron 
cyclotron frequency. The electric field 
\begin{eqnarray}
&\displaystyle
\tilde{\mathbf{E}}=\tilde{E}_{r}\mathbf{e}_{r}+\tilde{E}_{\varphi}\mathbf{e}_{\varphi}
+\tilde{E}\mathbf{e}_{z}=-\nabla \Phi \left(r,\varphi,z,t \right)
\label{2}
\end{eqnarray}
is the field of the electrostatic  plasma response on the helicon wave. 
The potential $\Phi \left(r,\varphi,z,t \right)$ is determined by the Poisson equation, 
\begin{eqnarray}
&\displaystyle 
-\bigtriangleup\Phi \left(r,\varphi,z,t \right)=
4\pi\sum_{\alpha=i,e} e_{\alpha}\int f_{\alpha}\left(\mathbf{v}, r,\varphi,z, t \right)d{\bf v}, 
\label{3}
\end{eqnarray}
in which $f_{\alpha}$ is the fluctuating part of the distribution function 
$F_{\alpha}$, $f_{\alpha}=F_{\alpha}-F_{0\alpha}$ and $F_{0\alpha}$ is the equilibrium 
distribution function.

\begin{figure}[ht]
\includegraphics[width=0.48\textwidth]{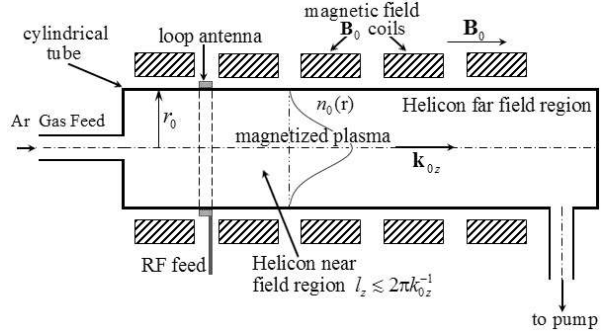}
\caption{\label{fig1} Schematic diagram of the helicon plasma source}
\end{figure}

In this paper, we consider the far-field region of the helicon wave, where the helicon 
wave is the travelling wave along the magnetic field direction. The electric, 
$\mathbf{E}_{1}\left(r, z, t \right)$, and magnetic, $\mathbf{B}_{1}\left(r,z,t \right)$, 
fields of the helicon wave are given in this region by the relations 
\begin{eqnarray}
&\displaystyle
E_{1r}\left(r,z,t \right)=E_{1r}\left(r \right) \sin\left(k_{0z}z-\omega_{0}t \right) ,
\nonumber
\\  
&\displaystyle
E_{1\varphi}\left(r,z,t \right)=E_{1\varphi}\left(r \right) \cos\left(k_{0z}z-\omega_{0}t \right),
\nonumber
\\  
&\displaystyle 
E_{1z}\left(r,z,t \right)=0,
\label{4}
\end{eqnarray}
where $E_{1r}$ is the radial, $E_{1\varphi}$ is the azimuthal components of the 
electric field $\mathbf{E}_{1}\left(r,\varphi,z,t \right)$ of the helicon wave, 
$\omega_{0}$ is the frequency and $k_{0z}$ is the wavenumber component of the helicon 
wave directed along magnetic field $\mathbf{B}_{0}=B_{0}\mathbf{e}_{z}$ ; 
\begin{eqnarray}
&\displaystyle
B_{1r}\left(r,z,t \right)=B_{1r}\left(r \right) \cos\left(k_{0z}z-\omega_{0}t \right),
\nonumber
\\  
&\displaystyle
B_{1\varphi}\left(r,z,t \right)=B_{1\varphi}\left(r \right) \sin\left(k_{0z}z-\omega_{0}t \right),
\nonumber
\\  
&\displaystyle
B_{1z}\left(r,z,t \right)=B_{1z}\left(r \right)  \sin\left(k_{0z}z-\omega_{0}t \right),
\label{5}
\end{eqnarray}
are the components of the magnetic field of the helicon wave. Note, that because magnetic 
field $\mathbf{B}_{1}$ satisfies the Gauss law $\nabla\cdot\mathbf{B}_{1}= 0$, the relation   
\begin{eqnarray}
&\displaystyle
\frac{1}{r}\frac{\partial}{\partial r}\left(rB_{1r}\left(r\right)\right)
+k_{0z}B_{1z}\left(r\right)=0
\label{6}
\end{eqnarray}
for ${B}_{1r}\left(r \right)$ and ${B}_{1z}\left(r \right)$ functions in Eq. (\ref{5}) occurs. 

The helicon wave field under the loop antenna and in its near-field, which is limited by 
the small distance 
$|l_{z}|\sim k^{-1}_{0z}$, has much more complicate spatial distribution along 
magnetic field and contains the propagating and the
exponentially decaying modes\cite{Gushchin}. In this narrow near-field zone, 
where the helicon wave electric field 
has the maximum values of the magnitude and has maximum gradient 
along the magnetic field, 
the strong ponderomotive force along the magnetic field\cite{Karpman} develops. 
This ponderomotive force determines the specific nonlinear processes of the helicon wave-plasma 
interactions, which are not investigated yet, that are different from the processes 
in the far-field region considered in this paper.

In the helicon wave field (\ref{4}), (\ref{5}), a plasma in equilibrium has an azimuthally 
symmetric radially inhomogeneous density profile. The radial profiles of the electric 
and magnetic fields depends on the plasma density profiles and differ from the profiles
\begin{eqnarray}
&\displaystyle
E_{1r}\left(r \right)=E_{1r}J_{1}\left(k_{0\perp}r \right), 
\nonumber
\\  
&\displaystyle
E_{1\varphi}\left(r\right)=E_{1\varphi}J_{1}\left(k_{0\perp}r \right), 
\nonumber
\\  
&\displaystyle
B_{1r}\left(r\right)=B_{1r}J_{1}\left(k_{0\perp}r \right),
\nonumber
\\  
&\displaystyle
B_{1\varphi}\left(r\right)=B_{1\varphi}J_{1}\left(k_{0\perp}r \right),
\nonumber
\\  
&\displaystyle
B_{1z}\left(r\right)=B_{1z}J_{0}\left(k_{0\perp}r \right),
\label{7}
\end{eqnarray}
where $k_{0\perp}=\omega_{0}\omega_{pe}^{2}/\left(\omega_{ce}k_{0z}c^{2}\right)$, 
and $J_{0,1}\left(k_{0\perp}r\right) $ are the Bessel functions, known for the 
helicon plasmas with a uniform density\cite{Chen2}. 
The simple relations 
\begin{eqnarray}
&\displaystyle
B_{1\varphi}=\frac{ck_{0z}}{\omega_{0}}E_{1r}, \quad  B_{1r}=-\frac{ck_{0z}}
{\omega_{0}}E_{1\varphi},
\label{8}
\end{eqnarray}
which stem from the Faradey's law, reveal that the $B_{1r}$ and  $B_{1\varphi}$ 
contained terms of the Lorentz force in Eq. (\ref{1}), 
\begin{eqnarray}
&\displaystyle
\left |-\frac{v_{z}}{c}B_{1\varphi}\right |=\left |-\frac{v_{z}k_{0z}}{\omega_{0}}
E_{1r}\right|\ll \left|E_{1r}\right|
\label{9}
\end{eqnarray}
and
\begin{eqnarray}
&\displaystyle
\left |-\frac{v_{z}}{c}B_{1r}\right |=\left |-\frac{v_{z}k_{0z}}{\omega_{0}}E_{1\varphi}\right|\ll 
\left|E_{1\varphi}\right|
\label{10}
\end{eqnarray}
are much less than the helicon electric field force because $|\omega_{0}|\gg |k_{0z}v_{Te}|$ 
for the helicon wave, and may be neglected in the Vlasov equation (\ref{1}). Without 
these terms, the simplified Vlasov equation with 
electron velocity $\mathbf{v}=\left(v_{\bot}, \phi, v_{z}\right)$, 
determined in the polar coordinates by the relations (see Fig. 1)
\begin{eqnarray}
&\displaystyle
v_{r}=v_{\bot}\cos \phi, \qquad v_{\varphi}=v_{\bot}\sin \phi,
\label{11}
\end{eqnarray}
where $v_{\bot}=\left(v_{r}^{2}+v_{\varphi}^{2}\right)^{1/2}$, and $\phi=\tan^{-1}
\left(v_{\varphi}/v_{r}\right)$ is the gyroangle, has a form
\begin{eqnarray}
&\displaystyle
\frac{\partial F_{e}}{\partial t}+v_{\perp}\cos\phi\frac{\partial F_{e}}{\partial r}
+\frac{v_{\perp}}{r}\sin\phi\frac{\partial F_{e}}{\partial \varphi}
+v_{z}\frac{\partial F_{e}}{\partial z}
\nonumber
\\  
&\displaystyle
+\frac{e}{m_{e}}\left(\sin\phi E_{\varphi}+\cos\phi E_{r} \right) \frac{\partial F_{e}}{\partial 
v_{\bot}}
\nonumber
\\  
&\displaystyle
-\left[\omega_{ce}+ \frac{v_{\perp}}{r}\sin\phi+\frac{e}{m_{e}v_{\bot}}\left(\sin\phi E_{r}-\cos\phi 
E_{\varphi}\right)\right.
\nonumber
\\  
&\displaystyle
\left.+\frac{eB_{1z}\left(r,z,t \right) }
{m_{e}c}\right]\frac{\partial F_{e}}{\partial \phi}
+\frac{e}{m_{e}}E_{z}\frac{\partial F_{e}}{\partial v_{z}}=0.
\label{12}
\end{eqnarray}
The solution of the Vlasov equation (\ref{12}) is presented in the next section. 

\section{The theory of the micro-instabilities of the inhomogeneous cylindrical plasma 
in the field of the azimuthally symmetric radially inhomogeneous helicon wave}\label{sec3} 

Helicon discharges contain two disparate spatial scales: the macroscale of the 
radial inhomogeneity of the helicon wave, which is commensurable with radial 
scale of the plasma density inhomogeneity, and the microscale, which is 
commensurable with the thermal Larmor radius of electrons, but is much less than the 
macroscale of the plasma and of the helicon wave inhomogeneities. In this section, 
we develop the two-scale approach to the analytical solution of the Vlasov equation (\ref{12}), 
in which the respond on both spatial scales of the cylindrical radially inhomogeneous plasma on the 
azimuthally symmetric radially inhomogeneous helicon wave is accounted for.
This two-scale approach is based on the employing of the cylindrical guiding center variables 
$R_{e}, \psi$, and of the electron Larmor orbit variables $\rho_{e}, \delta$, which are related 
to the original electron variables $r,\varphi, v_{\bot},\phi$ via 
\cite{Chibisov} 
\begin{eqnarray} 
&\displaystyle
R_{e}^{2}=\frac{1}{\omega_{ce}^2}\left(v^{2}_{\perp}+2v_{\perp}r\omega_{ce}\sin\phi+r^{2}
\omega_{ce}^{2} \right), 
\label{13}
\end{eqnarray}
\begin{eqnarray}
&\displaystyle
\psi=\varphi-\alpha, 
\label{14}
\end{eqnarray}
\begin{eqnarray}
&\displaystyle
\rho^{2}_{e}=\frac{v_{\perp}^2}{\omega_{ce}^{2}},
\label{15}
\end{eqnarray}
\begin{eqnarray}
&\displaystyle
\delta=\phi+\alpha,
\label{16}
\end{eqnarray}
\begin{eqnarray*}
&\displaystyle
\alpha=\arcsin\left[\frac{\cos\phi}{
\left(1+v^{-2}_{\perp}\left(r^{2}\omega_{ce}^{2}+2v_{\perp}r\omega_{ce}
\sin\phi \right)\right)^{1/2}}\right].
\end{eqnarray*}
The geometric interpretation of the cylindrical guiding center coordinates for an electron is 
presented in Fig. \ref{fig1}.  
In  coordinates $R_{e}$, $\psi$, $\rho_{e}$, $\delta$, $z$, $t$, Eq. (\ref{12}) 
transforms to the following equation for $F_{e}\left(R_{e},\psi, \rho_{e}, \delta, z, t\right)$:
\begin{eqnarray}
&\displaystyle
\frac{\partial F_{e}}{\partial t}+v_{z}\frac{\partial F_{e}}{\partial z}-\omega_{ce}\frac{\partial 
F_{e}}{\partial \delta}
\nonumber
\\  
&\displaystyle
+\frac{c}{B_{0}}\frac{1}{\left(R^{2}_{e}-2\rho_{e}R_{e}\sin \delta + \rho^{2}_{e}\right)^{1/2}}
\nonumber
\\  
&\displaystyle
\times
\left[\left(E_{1r}\rho_{e}\cos \delta  +
E_{1\varphi}\left(R_{e}-\rho_{e}\sin \delta\right)\right)\frac{\partial F_{e}}{\partial R_{e}}\right.
\nonumber
\\  
&\displaystyle
-\left.\left(\left(1-\frac{\rho_{e}}{R_{e}}\sin\delta\right)E_{1r}+\frac{\rho_{e}}{R_{e}}\cos \delta 
E_{1\varphi}\right)\frac{\partial F_{e}}{\partial\psi}\right.
\nonumber
\\  
&\displaystyle
\left.+\left(E_{1r}R_{e}\cos \delta +E_{1\varphi}\left(R_{e}\sin\delta -\rho_{e}\right)\right)
\frac{\partial F_{e}}{\partial \rho_{e}}
\right.
\nonumber
\\  
&\displaystyle
+\left(E_{1r}\left(2-\sin \delta \frac{\left(R^{2}_{e}+\rho^{2}_{e}\right)}{R_{e}
\rho_{e}}\right)\left. 
+E_{1\varphi}\cos \delta\frac{\left(R^{2}_{e}+\rho^{2}_{e}\right)}{R_{e}\rho_{e}}\right)
\frac{\partial 
F_{e}}{\partial\delta}\right]
\nonumber
\\  
&\displaystyle
-\frac{e}{m_{e}c}B_{1z}\left[\rho_{e}\cos \delta\frac{\partial F_{e}}{\partial R_{e}}
+ \frac{\partial F_{e}}{\partial \delta} 
-\sin \delta \frac{\rho_{e}}
{R_{e}}\left(\frac{\partial F_{e}}{\partial \delta}
-\frac{\partial F_{e}}{\partial \psi}\right)\right]
\nonumber
\\  
&\displaystyle
-\frac{c}{B_{0}R_{e}}\left(\frac{\partial \Phi}{\partial \psi}-\frac{\partial \Phi}{\partial \delta}
\right)\frac{\partial F_{e}}{\partial R_{e}}
+\frac{c}{B_{0}\rho_{e}}\frac{\partial \Phi}{\partial 
\delta}\frac{\partial F_{e}}{\partial\rho_{e}}
-\frac{e}{m_{e}}\frac{\partial \Phi}{\partial z}\frac{\partial F_{e}}{\partial v_{ez}}=0.
\label{17}
\end{eqnarray}
In the helicon discharge,  $\rho_{e}\ll R_{e}$, excluding small region $R_{e}\sim \rho_{e}$ of the 
discharge center. For example, for the magnetic field $B_{0}=50$\,mT and electron 
temperature $T_{e}=4$\,eV the thermal electron Larmor radius $\rho_{e}= 10^{-1}$\,cm, whereas the 
radial scales of the helicon wave and of a plasma inhomogeneities are typically a few centimetres.
Equation (\ref{17}), in which the terms on the order of $\rho_{e}/R_{e}\ll 1$ are omitted, becomes
\begin{eqnarray}
&\displaystyle
\frac{\partial F_{e}}{\partial t}+v_{z}\frac{\partial F_{e}}{\partial z}
+\frac{c}{B_{0}}E_{1\varphi}\frac{\partial F_{e}}{\partial R_{e}} - \frac{c}{B_{0}}E_{1r}
\frac{\partial F_{e}}{R_{e}\partial\psi}
\nonumber
\\  
&\displaystyle
+\frac{c}{B_{0}}\left(E_{1r}\cos \delta+E_{1\varphi}\sin \delta\right)\frac{\partial F_{e}}
{\partial \rho_{e}}
\nonumber
\\  
&\displaystyle
+\left(2\frac{c}{B_{0}R_{e}}E_{1r}-\omega_{ce} 
-\frac{c}{B_{0}\rho_{e}}\frac{1}{\rho_{e}}
\left(E_{1r}\sin \delta-E_{1\varphi}\cos \delta\right)\right)\frac{\partial F_{e}}{\partial\delta}
\nonumber
\\  
&\displaystyle
-\frac{e}{m_{e}c}B_{1z}\frac{\partial F_{e}}{\rho_{e}\partial\delta}
=\frac{c}{B_{0}R_{e}}\left(\frac{\partial \Phi}{\partial \psi}-\frac{\partial \Phi}{\partial \delta}
\right)\frac{\partial 
F_{e}}{\partial R_{e}}
\nonumber
\\  
&\displaystyle
-\frac{c}{B_{0}\rho_{e}}\frac{\partial \Phi}{\partial \delta}\frac{\partial 
F_{e}}{\partial\rho_{e}}
+\frac{e}{m_{e}}\frac{\partial \Phi}{\partial z}\frac{\partial F_{e}}{\partial v_{ez}}.
\label{18}
\end{eqnarray}
The Vlasov equation (\ref{18}) with $\Phi=0$ is the equation for the equilibrium 
electron distribution function $F_{e0}$. Consider now the system of equations for the 
characteristics of equation for $F_{e0}$,
\begin{eqnarray}
&\displaystyle
dt=\frac{dR_{e}}{\frac{c}{B_{0}}E_{1\varphi}}=\frac{d\psi}{-\frac{c}{B_{0}R_{e}}E_{1r}}=\frac{d
\rho_{e}}{\frac{c}
{B_{0}}\left(E_{1r}\cos \delta+E_{1\varphi}\sin\delta\right)}
\nonumber
\\  
&\displaystyle
=d\delta\left[ -\omega_{ce}+2\frac{c}{B_{0}R_{e}}E_{1r}-\frac{eB_{1z}}{cm_{e}}\right. 
\nonumber
\\  
&\displaystyle
\left. 
-\frac{c}{B_{0}\rho_{e}}\left(E_{1r}\sin 
\delta-E_{1\varphi} \cos \delta\right)\right] ^{-1}=\frac{dz}{v_{z}}.
\label{19}
\end{eqnarray}
With approximations $E_{1r}\left(r\right)\approx E_{1r}\left(R_{e}\right)$ and 
$E_{1\varphi}\left(r\right)\approx E_{1\varphi}\left(R_{e}\right)$, which follows from the relation
\begin{eqnarray}
&\displaystyle
r=\left(R^{2}_{e}-2\rho_{e}R_{e}\sin\delta+\rho^{2}_{e} \right)^{1/2}\approx R_{e}
\label{20}
\end{eqnarray} 
in the limit $\rho_{e}\ll R_{e}$, the system of equations for the macroscale guiding center coordinates 
$R_{e}$, $\psi$
\begin{eqnarray}
&\displaystyle
dt=\frac{dR_{e}}{\frac{c}{B_{0}}E_{1\varphi}\left(R_{e}\right)\cos \left(\omega_{0}t
-k_{0z}\left(z_{1}+v_{z}t\right)\right)}
\nonumber
\\  
&\displaystyle
=\frac{d\psi}{-\frac{c}{B_{0}R_{e}}E_{1r}\left(R_{e}\right)\sin \left(\omega_{0}t-k_{0z}\left(z_{1}
+v_{z}t\right)\right)},
\label{21}
\end{eqnarray}
where $z_{1}=z-v_{z}t$ is the integral of system (\ref{19}), becomes separate from the system 
of equation for the microscale coordinates $\rho_{e}$ and $\delta$ of the Larmor motion. By direct integration 
of Eq. (\ref{21}) for $R_{e}\left(t\right)$ and for  $\psi\left(t\right)$ we derive the equations 
\begin{eqnarray}
&\displaystyle
R_{e1}=R_{e}\left(t\right)-\frac{c}{B_{0}}\int dt_{1}E_{1\varphi}\left(R_{e}\left(t_{1}\right)\right)
\cos \left(\omega_{0}t_{1}-k_{0z}\left(z_{1}+v_{z}t_{1}\right)\right)
\label{22}
\end{eqnarray}
and
\begin{eqnarray}
&\displaystyle 
\psi_{1}=\psi\left(t\right)+\frac{c}{B_{0}}\int dt_{1}\frac{1}{R_{e}\left(t_{1}\right)}E_{1r}
\left(R_{e}\left(t_{1}\right)\right)\sin 
\left(\omega_{0}t_{1}-k_{0z}\left(z_{1}+v_{z}t_{1}\right)\right),
\label{23}
\end{eqnarray}
where $R_{e1}$ and $\psi_{1}$ are the integrals of system (\ref{21}). By employing the method of 
successive approximation we derive the approximate solution of Eq. (\ref{22}) 
for $R_{e}\left(t\right)$ in the form 
\begin{eqnarray}
&\displaystyle
R_{e}\left(t \right) \approx R_{e1}+\frac{c}{B_{0}\omega_{0}}E_{1\varphi}\left(R_{e1} \right)
\sin\left( \omega_{0}t-k_{0z}z_{1}\right) 
\nonumber
\\  
&\displaystyle
-\frac{c^{2}}{B_{0}^{2}\omega_{0}^{2}}\frac{1}{4}\sin^{2}\left( \omega_{0}t-k_{0z}z_{1}\right) 
\frac{d}{dR_{e1}}E_{1\varphi}^{2}\left(R_{e1} \right),
\label{24}
\end{eqnarray}
which is the  power series expansion in $\left|\xi/R_{e1}\right|\ll 1$, where $\xi=\frac{c}{B_{0}
\omega_{0}}E_{1\varphi}\left(R_{e1} \right)$ 
is the amplitude of the displacement of an electron along the coordinate $R_{e}$ at 
$R_{e}=R_{e1}$. In this solution,
the term $k_{0z}v_{z}t$ is omitted, because for the helicon wave $\omega_{0}\gg k_{0z}v_{Te}$, 
where $v_{Te}$ is the electron thermal velocity.

\begin{figure}[ht]
\includegraphics[width=0.4\textwidth]{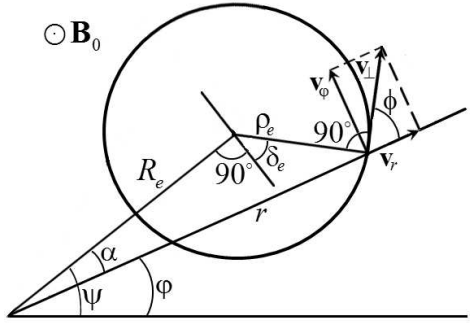}
\caption{\label{fig2} The geometric interpretation of the cylindrical guiding center 
coordinates for an electron.}
\end{figure}

The approximate solution to Eq. (\ref{23}) for the angle $\psi$
in the form of the power series expansion in $\left|\xi/R_{e1}\right|\ll 1$ with accounting 
for the terms on the first and the second order of $\left|\xi/R_{e1}\right|^{2}$ has a form
\begin{eqnarray}
&\displaystyle
\psi \approx \psi_{1}-\frac{c}{B_{0}R_{e1}\omega_{0}}E_{1r}\left(R_{e1} \right) 
\cos\left( \omega_{0}t-k_{z0}z_{1}\right)
\nonumber
\\  
&\displaystyle
-\frac{c^{2}E_{1\varphi}\left(R_{e1} \right)}{2B_{0}^{2}R_{e1}^{2}\omega_{0}}
\left(E_{1r}\left(R_{e1} \right)-R_{e1}\frac{\partial E_{1r}\left(R_{e1} \right)}{\partial R_{e1}}
\right) 
\nonumber
\\  
&\displaystyle
\times
\left[ t-\frac{1}{2\omega_{0}}\sin2\left( \omega_{0}t-k_{z0}z_{1}\right)
\right]. 
\label{25}
\end{eqnarray}
It contains the terms oscillating on frequencies $\omega_{0}$ and $2\omega_{0}$, and the term 
corresponding to the rotation of the guiding 
center coordinate with stationary radially inhomogeneous angular velocity $\Omega_{e}\left(R_{e1}
\right)$,
\begin{eqnarray}
&\displaystyle
\Omega_{e}\left(R_{e1}\right) = \frac{c^{2}E_{1\varphi}\left(R_{e1}
\right)}{2B_{0}^{2}R_{e1}^{2}\omega_{0}}\left[ E_{1r}\left(R_{e1} \right)
-R_{e1}\frac{\partial E_{1r}\left(R_{e1} \right)}{\partial R_{e1}}
\right]. 
\label{26}
\end{eqnarray}
The discovered effect of the electron component rotation with angular velocity $\Omega_{e}
\left(R_{e1}\right)$ in the cylindrical azimuthally
symmetric $\left(m_{0}=0\right)$ helicon wave is the first result 
of our two-scales analysis of the Vlasov equation solution.This effect is missed in the 
slab model of plasma in the spatially uniform helicon wave field. The order on value estimate for 
$B_{0}=50$\,mT, $\omega_{0}=10^{7}$\,s$^{-1}$, $E_{1r}=E_{1\varphi}=5$\,V/cm,  and $R_{e1}=2.5$\,cm
gives $\Omega_{e}=8\cdot 10^{5}$\,s$^{-1}$.
In what follows, we use the approximation 
\begin{eqnarray}
&\displaystyle
\psi=\psi_{1}-a_{e}\left(R_{e1}\right)\cos\left(\omega_{0}t-k_{0z}z_{1}\right)-\Omega_{e}\left(R_{e1}
\right)t,
\label{27}
\end{eqnarray}
where 
\begin{eqnarray}
&\displaystyle
a_{e}\left(R_{e1}\right)=\frac{cE_{1r}\left(R_{e1} \right)}{B_{0}R_{e1}\omega_{0}},
\label{28}
\end{eqnarray}
which is valid for a time $t\sim \Omega_{0}^{-1}\gg \omega_{0}^{-1}$. 

Because $\omega_{ce}$ is much larger than any other term in the equation for $d\delta/dt$ 
of system (\ref{19}), the solution for $\delta_{1}$  is determined with a great accuracy as 
\begin{eqnarray}
&\displaystyle
\delta=\delta_{1}-\omega_{ce}t.
\label{29}
\end{eqnarray}
The solution of the equation for the radius $\rho_{e}$ of the electron Larmor orbit,
\begin{eqnarray}
&\displaystyle
\frac{d\rho_{e}}{dt}=\frac{c}{B_{0}}\Big[E_{1r}\left(R_{e} \right)
\sin \left(k_{z0}z_{1}-\omega_{0}t\right)\cos\left(\delta_{1}-\omega_{ce}t \right) 
\nonumber
\\  
&\displaystyle
+E_{1\varphi}\left(R_{e} \right)
\cos \left(\omega_{0}t-k_{z0}z_{1}\right)\sin\left(\delta_{1}-\omega_{ce}t \right) 
\Big],
\label{30}
\end{eqnarray}
where $\delta_{1}$ is given by Eq. (\ref{29}), is
\begin{eqnarray}
&\displaystyle
\rho_{e}=\rho_{e1}-\frac{c}{B_{0}\omega_{ce}}
\Big[E_{1r}\left(R_{e1} \right)
\sin \left(\omega_{0}t-k_{z0}z_{1}\right)
\sin\left(\omega_{ce}t-\delta_{1} \right) 
\nonumber
\\  
&\displaystyle
-E_{1\varphi}\left(R_{e1} \right)
\cos \left(\omega_{0}t-k_{z0}z_{1}\right)\cos\left(\omega_{ce}t-\delta_{1} \right) 
\Big] 
\label{31}
\end{eqnarray}
with accuracy to terms of the order of $O\left(\frac{\omega_{0}}{\omega_{ce}}\ll 1 \right)$. 

It is easy to check that the Vlasov equation for the equilibrium electron distribution function 
$F_{e0}$ in variables $R_{e1}, \psi_{1}, \rho_{e1}, \delta_{1}, v_{z}, z_{1}$ determined by the 
solutions (\ref{24}), (\ref{25}), (\ref{31}) and (\ref{29}), respectively, reduces to the equation 
\begin{eqnarray*}
&\displaystyle
\frac{\partial F_{e0}}{\partial t}=0,
\end{eqnarray*}
and, therefore, $F_{e0}= F_{e0}\left(R_{e1}, \psi_{1}, \rho_{e1}, \delta_{1}, v_{z}, z_{1}\right)$, 
and does not depend on time variable. The equation for the perturbation $f_{e}\left(R_{e1}, \psi_{1}, 
\rho_{e1}, \delta_{1}, v_{z}, z_{1}, t\right)$ of the equilibrium distribution function 
$F_{e0}\left(R_{e1}, \rho_{e1}, v_{z} \right)$  of the azimuthally symmetric radially inhomogeneous 
plasma,
\begin{eqnarray}
&\displaystyle
\frac{d}{dt}f_{e}\left(R_{e1}, \psi_{1}, \rho_{e1}, 
\delta_{1}, v_{z}, z_{1}, t\right)
=\left(\frac{c}{B_{0}R_{e1}}\left(\frac{\partial \Phi}{\partial \psi_{1}}
- \frac{\partial\Phi}{\partial \delta_{1}}
\right)	\frac{\partial }{\partial R_{e1}}\right.
\nonumber
\\  
&\displaystyle
\left.-\frac{c}{B_{0}}\frac{1}{\rho_{e1}}\frac{\partial\Phi}{\partial \delta_{1}}
\frac{\partial}{\partial \rho_{e1}} 
+\frac{e}{m_{e}}\frac{\partial\Phi}{\partial z_{1}}
\frac{\partial }{\partial v_{ez}}\right)
F_{e0}\left(R_{e1}, \rho_{e1}, v_{z} \right). 
\label{32}
\end{eqnarray}
follows from Eq. (\ref{18}) in which variables $R_{e}, \psi, \rho_{e}, 
\delta, v_{z}, z, t$ are transformed on \\ $R_{e1}, \psi_{1}, \rho_{e1}, 
\delta_{1}, v_{z}, z_{1}, t$, by employing the relations (\ref{24}), (\ref{25}), 
(\ref{27}), (\ref{29}), (\ref{31}).
In this equation, potential $\Phi\left(R_{e1}, \psi_{1}, \rho_{e1},\delta_{1}, 
v_{z},z_{1},t\right)$ 
is presented in the form of the Fourier-Bessel transformation,
\begin{eqnarray}
&\displaystyle
\Phi\left(R_{e1}, \psi_{1}, \rho_{e1},\delta_{1}, v_{z},z_{1},t\right)
=\frac{1}{\left(2\pi\right)^{4}}\sum\limits^{\infty}_{m=-\infty}\sum\limits^{\infty}_{n=-\infty}
\nonumber
\\  
&\displaystyle
\times
\int
dk_{\perp}k_{\perp}dk_{z}d\theta d\omega \Phi\left(k_{\perp}, \theta, k_{z}, \omega \right)
J_{n}\left(k_{\perp}\rho_{e1} \right) J_{n+m}\left(k_{\perp}R_{e1} \right) 
\nonumber
\\  
&\displaystyle
\times
\exp\left[-in\left(\delta_{1}-\omega_{ce}t\right)-im\left(\theta-\psi_{1}
+a_{e}\left(R_{e1}\right)\cos\left(\omega_{0}t-k_{0z}z_{1}\right)+\Omega_{e}\left(R_{e1}\right)t \right)\right.
\nonumber
\\  
&\displaystyle
\left.+i\left(m+n \right) \frac{\pi}{2}-i\left(\omega-k_{z}v_{z}\right)t+ik_{z}z_{1} \right]. 
\label{33}
\end{eqnarray}
It was derived from the Fourier-Bessel transform 
\begin{eqnarray}
&\displaystyle
\Phi\left(r,\varphi,z,t \right) =\frac{1}{\left(2\pi\right)^{4}}\int \Phi\left(\mathbf{k},\omega \right) 
e^{-i\omega t+ik_{\perp}r\cos\left(\theta-\varphi\right)+ik_{z}z }k_{\perp}dk_{\perp}d\theta dk_{z}d\omega ,
\label{34}
\end{eqnarray}
in which the identity
\begin{eqnarray}
&\displaystyle
k_{\perp}r\cos\left(\theta-\varphi \right) =k_{\perp}R_{e}\cos\left(\theta-\psi \right) 
+k_{\perp}\rho_{e}\sin\left(\theta-\psi-\delta \right) 
\label{35}
\end{eqnarray}
and Eqs. (\ref{27}) and (\ref{29}) were employed. The solution for $f_{e}\left(R_{e1}, 
\psi_{1}, \rho_{e1}, \delta_{1}, v_{z}, z_{1}, t\right)$ of Eq. (\ref{32}) with potential 
(\ref{33}) is
\begin{eqnarray}
&\displaystyle
f_{e}\left(R_{e1}, \psi_{1}, \rho_{e1}, \delta_{1}, v_{z}, z_{1}, t\right)=-\frac{1}{\left(2\pi\right)^{4}}
\sum\limits_{m=-\infty}^{\infty}
\sum\limits_{n=-\infty}^{\infty}\sum\limits_{p=-\infty}^{\infty}\int dk_{\bot}k_{\bot}d\theta dk_{z}d\omega
\Phi\left(\mathbf{k},\omega \right)
\nonumber
\\  
&\displaystyle
\times\frac{J_{n}\left(k_{\bot}\rho_{e1}\right)J_{m+n}\left(k_{\bot}R_{e1}\right)J_{p}\left(ma_{e}\left(R_{e1}\right)\right)}
{\omega-k_{z}v_{z}-n\omega_{ce}-p\omega_{0}+m\Omega_{e}\left(R_{e1}\right)}
\nonumber
\\  
&\displaystyle
\times
\left[\frac{c}{B_{0}}\left(\left(m+n\right)\frac{1}{R_{e1}}
\frac{\partial F_{e0}}{\partial R_{e1}}+n\frac{1}{\rho_{e1}}\frac{\partial F_{e0}}{\partial \rho_{e1}}\right)+\frac{e}{m_{e}}k_{z}
\frac{\partial F_{e0}}{\partial v_{ez}}\right]
\nonumber
\\  
&\displaystyle
\times
\exp\left(-i\left(\omega-k_{z}v_{z}-n\omega_{ce}-p\omega_{0}+m\Omega_{e}\left(R_{e1}\right)\right)t
-in\delta_{1}-im\left(\theta-\psi_{1}\right)\right.
\nonumber
\\  
&\displaystyle
\left.+i\left(k_{z}-pk_{0z}\right)z_{1}+i\left(m+n-p\right)\frac{\pi}{2}\right).
\label{36}
\end{eqnarray}

The dynamics of ions in the helicon wave is different from the electron dynamics. 
The ion cyclotron frequency, 
$\omega_{ci}$, is much less than the frequency of the helicon wave $\omega_{0}$. 
(For the magnetic field $B_{0}=50$\,mT 
and argon gas $\omega_{ci}=1.25\cdot 10^{4}$s$^{-1}$ $\ll \omega_{0}=10^{7}$s$^{-1}$ ). 
Therefore, the ions displacement 
in the helicon wave is as of the unmagnetized particle, and is estimated by 
$\delta r_{i}\sim e_{i}E_{1}/m_{i}\omega^{2}_{0}\sim 10^{-3}$cm 
for $E_{1}=5$\,V/cm. This displacement is much less than the thermal argon ion Larmor radius 
$\rho_{i}$, which for 
$T_{i}=2.6\cdot 10$ $^{-2}$ eV and $B_{0}=50$\,mT  is on the order of 2\,cm. Therefore, 
the thermal motion of ions is practically unaffected by the helicon wave. 

By using the solution (\ref{36}) for $f_{e}$ in the Poisson equation (\ref{3}), we 
derive the integral equation for the $m$-th harmonic $\Phi_{m}\left(k_{\bot},k_{z},\omega\right)$ 
of the Fourier-Bessel transformed potential determined by the relation
\begin{eqnarray}
&\displaystyle
\Phi_{m}\left(k_{\bot},k_{z},\omega\right) =\frac{1}{2\pi}\int d\theta_{1}\Phi\left(k_{\bot},
\theta_{1},k_{z},\omega \right)e^{-im\theta_{1}}.
\label{37}
\end{eqnarray}
This equation has a form
\begin{eqnarray}
&\displaystyle
\Phi_{m}\left(k_{\bot},k_{z},\omega\right)\left(1-\frac{\omega^{2}_{pi}}{\omega^{2}}\right)
\nonumber
\\  
&\displaystyle
+8\pi^{2}\frac{e^{2}}{k^{2}m_{e}}\omega^{2}_{ce}
\sum^{\infty}_{n=-\infty}\sum^{\infty}_{p=-\infty}\sum^{\infty}_{p_{1}=-\infty}\int\limits^{\infty}_{0}
dR_{e1}R_{e1}\int\limits^{\infty}_{0}d\rho_{e1}\rho_{e1}
\int \limits^{\infty}_{-\infty}dv_{z}
\nonumber
\\  
&\displaystyle
\times \int\limits_{0}^{\infty}dk_{1\bot}k_{1\bot}\Phi_{m}\left(k_{1\bot},k_{z}-\left(p-p_{1}\right)k_{0z},
\omega-\left(p-p_{1}\right)\omega_{0}\right)e^{i\left(p-p_{1}\right)\frac{\pi}{2}}
\nonumber
\\  
&\displaystyle
\times\frac{J_{n}\left(k_{\bot}\rho_{e1}\right)J_{n}\left(k_{1\bot}\rho_{e1}\right)J_{n+m}
\left(k_{\bot}R_{e1}\right)J_{n+m}\left(k_{1\bot}R_{e1}\right)J_{p}\left(ma_{e}\left(R_{e1}\right)\right)J_{p_{1}}\left(ma_{e}\left(R_{e1}\right)\right)}
{\omega-n\omega_{ce}-p\omega_{0}-k_{z}v_{z}+m\Omega_{e}\left(R_{e1}\right)}
\nonumber
\\  
&\displaystyle
\times \left[\frac{\left(m+n\right)}{\omega_{ce}R_{e1}}\frac{\partial F_{e0}}{\partial 
R_{e1}}+\frac{n}{\omega_{ce}\rho_{e1}}\frac{\partial F_{e0}}{\partial \rho_{e1}}+k_{z}
\frac{\partial F_{e0}}{\partial v_{z}}\right]=0.
\label{38}
\end{eqnarray}
In Eq. (38), the approximation of the unmagnetized ions was used, which is 
applicable for the treating of the 
instabilities with the growth rate $\gamma\left(\mathbf{k}\right)\gg \omega_{ci}/2\pi$ 
and $k_{\bot}\rho_{i}\gg 1$. The solution to Eq. (\ref{38}) for 
$\Phi_{m}\left(k_{\bot},k_{z},\omega\right)$ is the cylindrical 
wave with a continuous spectrum characterized by the wave number $k_{\bot}$ and 
azimuthal number $m$. Here, we derive the solution to Eq. (\ref{38}) in the short 
wavelength limit
\begin{eqnarray}
&\displaystyle
k_{\bot}R_{e1}\sim m\gg 1, 
\label{39}
\end{eqnarray}
employing the approach developed in Ref. \cite{Mikhailenko2} in the studies of the 
drift turbulence of the azimuthally symmetric radially nonuniform plasma and 
applied in Ref.\cite{Mikhailenko3} in the studies 
of the shear flow driven ion cyclotron and ion acoustic instabilities of the cylindrical 
inhomogeneous plasma.  The integration over macroscale $R_{e1}$ and over microscale 
wave number $k_{1\bot}$, performed in  Ref.\cite{Mikhailenko2}, 
reveal that the  vicinity of the $R_{e1}=\frac{m}{k_{1\bot}}=R_{e0}$ values 
and the vicinity of the $k_{1\bot}= k_{\bot}$ values give the dominant input to 
the integrals over $R_{e1}$  and $k_{1\bot}$ in Eq. (\ref{38}). It is important to note, 
that the region of $R_{e1}\approx R_{e0}$ 
corresponds approximately to the region of the first maximum of the 
$J_{m}\left(k_{\bot}R_{e1}\right)$ Bessel function, 
where the known cosine asymptotic, which is valid for $k_{\bot}R_{e1}\gg m$, is not applicable, and 
Eq. (\ref{38}) can not be approximated by the plane geometry model. After the integration 
over $R_{e1}$ and $k_{1\bot}$, Eq. (\ref{38}) becomes
\begin{eqnarray}
&\displaystyle
\Phi_{m}\left(k_{\bot},k_{z},\omega\right)\left(1-\frac{\omega^{2}_{pi}}{\omega^{2}}\right)
\nonumber
\\  
&\displaystyle
+8\pi^{2}\frac{e^{2}}{k^{2}m_{e}}\omega^{2}_{ce}
\sum^{\infty}_{n=-\infty}\sum^{\infty}_{p=-\infty}\sum^{\infty}_{p_{1}=-\infty}
\int\limits^{\infty}_{0}d\rho_{e1}\rho_{e1}
\int \limits^{\infty}_{-\infty}dv_{z}
\nonumber
\\  
&\displaystyle
\times \Phi_{m}\left(k_{\bot},k_{z}-\left(p-p_{1}\right)k_{0z},
\omega-\left(p-p_{1}\right)\omega_{0}\right)e^{i\left(p-p_{1}\right)\frac{\pi}{2}}
\nonumber
\\  
&\displaystyle
\times\frac{J^{2}_{n}\left(k_{\bot}\rho_{e1}\right)
J_{p}\left(ma_{e}\left(R_{e0}\right)\right)J_{p_{1}}\left(ma_{e}\left(R_{e0}\right)\right)}
{\omega-n\omega_{ce}-p\omega_{0}-k_{z}v_{z}+m\Omega_{e}\left(R_{e0}\right)}
\nonumber
\\  
&\displaystyle
\times \left[\frac{\left(m+n\right)}{\omega_{ce}R_{e0}}\frac{\partial F_{e0}}{\partial 
R_{e0}}+\frac{n}{\omega_{ce}\rho_{e1}}\frac{\partial F_{e0}}{\partial \rho_{e1}}+k_{z}
\frac{\partial F_{e0}}{\partial v_{z}}\right]=0.
\label{40}
\end{eqnarray}
The cylindrical plasma excited by the cylindrical azimuthally
symmetric $\left(m_{0}=0\right)$ helicon wave has an azimuthally symmetric 
radially inhomogeneous density and temperature profiles. We will consider in 
what follows the equilibrium distribution function $F_{e0}
\left(\rho_{e1}, v_{z}, R_{e0}\right)$ 
for Eq. (\ref{40}) as a Maxwellian
\begin{eqnarray}
&\displaystyle
F_{e0}\left(\rho_{e1}, v_{z}, R_{e0}\right)=\frac{n_{e0}\left(R_{e0}\right)}{\left(2\pi\right)^{3/2}
v^{3}_{Te}}\exp\left[-\frac{\rho^{2}_{e1}}{2\rho^{2}_{Te}}-\frac{v^{2}_{z}}{v^{2}_{Te}}\right] , 
\label{41}
\end{eqnarray}
where $\rho_{Te}=v_{Te}\left(R_{e0}\right)/\omega_{ce}$, and $v^{2}_{Te}\left(R_{e0}\right)
=T_{e}\left(R_{e0}\right)/m_{e}$. After integration of Eq. (\ref{40}) over $\rho_{e1}$ and $v_{z}$ 
with Maxwellian distribution (\ref{41}), we derive the basic equation for $\Phi_{m}
\left(k_{\bot},k_{z},\omega\right)$, 
\begin{eqnarray}
&\displaystyle
\Phi_{m}\left(k_{\bot},k_{z},\omega\right)\left(1-\frac{\omega^{2}_{pi}}{\omega^{2}}\right)
\nonumber
\\  
&\displaystyle
+\sum^{\infty}_{p=-\infty}\sum^{\infty}_{p_{1}=-\infty}J_{p}\left(ma_{e}\left(R_{e0}\right)\right)
J_{p_{1}}\left(ma_{e}\left(R_{e0}\right)\right)
\nonumber
\\  
&\displaystyle
\times\Phi_{m}\left(k_{\bot},k_{z}-\left(p-p_{1}\right)k_{0z}, 
\omega-\left(p-p_{1}\right)\omega_{0}\right)e^{i\left(p-p_{1}\right)\frac{\pi}{2}}
\nonumber
\\  
&\displaystyle
\times \varepsilon_{m(e)}\left(k_{\bot}, k_{z}, \omega-p\omega_{0}\right)=0,
\label{42}
\end{eqnarray}
where
\begin{eqnarray}
&\displaystyle
\varepsilon_{m(e)}\left(k_{\bot}, k_{z}, \omega-p\omega_{0}\right)
\nonumber
\\  
&\displaystyle
=\frac{1}{k^{2}\lambda^{2}_{De}}
\left\{1+i\sqrt{\pi}\sum^{\infty}_{n=-\infty}\left(\omega+m\Omega_{e}\left(R_{e0}\right)-p
\omega_{0}-\left(m+n\right)
\omega_{de}\left(R_{e0}\right)\left(1-\frac{1}{2}\eta_{e}\right)\right)
\right.
\nonumber
\\  
&\displaystyle
\left. \times \frac{1}{\sqrt{2}k_{z}v_{Te}}W\left(z_{en,p}\right)I_{n}\left(k_{\bot}^{2}\rho^{2}_{e}
\right)e^{-k_{\bot}^{2}\rho^{2}_{e}}\right.
\nonumber
\\  
&\displaystyle
\left.
-\eta_{e}\sum^{\infty}_{n=-\infty}\frac{\left(m+n\right)\omega_{de}\left(R_{e0}\right)}
{\sqrt{2}k_{z}v_{Te}}e^{-k_{\bot}^{2}\rho^{2}_{e}}\left[z_{en,p}\left(1+i\sqrt{\pi}z_{en,p}
W\left(z_{en,p}\right)\right)I_{n}\left(k_{\bot}^{2}\rho^{2}_{e}\right)
\right.\right.
\nonumber
\\  
&\displaystyle
\left.\left.-i\sqrt{\pi}W\left(z_{en,p}\right)k_{\bot}^{2}\rho^{2}_{e}\left(I_{n}\left(k_{\bot}^{2}
\rho^{2}_{e}\right)
-I'_{n}\left(k_{\bot}^{2}\rho^{2}_{e}\right)\right)\right]\right\}=0.
\label{43}
\end{eqnarray}
In Eq. (\ref{43}), $\lambda_{De}$ is the electron Debye length, $W\left(z_{e}\right)
=e^{- z_{e}^{2}}\left(1 +\left(2i / \sqrt {\pi 
}\right)\int\limits_{0}^{z_{e}} e^{t^{2}}dt \right)$ is the Faddeeva function\cite{Faddeyeva} 
with argument $z_{e}$ equal to
\begin{eqnarray}
&\displaystyle
z_{e}=\frac{\omega-p\omega_{0}+m\Omega_{e}\left(R_{e0}\right)-n\omega_{ce}}{\sqrt{2}k_{z}v_{Te}}, 
\label{44}
\end{eqnarray}
$I_{n}$ is the modified Bessel function of the first kind and order $n$, the prime in $I'_{n}$ 
denotes the derivative
with respect to the argument $k^{2}_{\bot}\rho^{2}_{Te}$ of the $I_{n}$ function, 
$\omega_{de}\left(R_{e0}\right)$ is the local 
electron diamagnetic drift frequency 
\begin{eqnarray}
&\displaystyle
\omega_{de}\left(R_{e0}\right)=\omega_{ce}\rho^{2}_{Te}\frac{\partial \ln n_{0e}\left(R_{e0}
\right)}{R_{e0}\partial R_{e0}}, 
\label{45}
\end{eqnarray}
and $\eta_{e}=\partial \ln T_{e}/\partial \ln n_{e0}$.

Equation (\ref{42}) is the basic equation, which determines in the short wavelength limit 
(\ref{39}) the electrostatic respond of the cylindrical radially inhomogeneous 
plasma on the cylindrical helicon wave. 
Equation (\ref{42}) is in fact the infinite system of equations 
for the potential $\Phi_{m}
\left(k_{\bot},k_{z},\omega\right)$ and for the infinite number of satellites \\
$\Phi_{m}\left(k_{1\bot},k_{z}-\left(p-p_{1}\right)k_{0z}, 
\omega-\left(p-p_{1}\right)\omega_{0}\right)$ \\ of this potential, which are derived from 
Eq. (\ref{42}) by changing in Eq. (\ref{42}) $\omega$ on $\omega-\left(p-p_{1}\right)\omega_{0}$ 
and $k_{z}$ on $k_{z}-\left(p-p_{1}\right)k_{0z}$, where $p$ and $p_{1}$ are the integral numbers.
The equality to zero of the determinant of this homogeneous system gives the general dispersion 
equation the solution of which determines the dispersive properties of the parametric 
instabilities. 

It was found analytically for the slab plasma geometry\cite{Akhiezer, Aliev}, that the 
relative oscillating motion of electrons and ions under the action of the helicon wave 
with frequency $\omega_{0}$ is a source of the development of the instabilities of the 
parametric type. The motion of electrons in the field of the cylindrical radially 
inhomogeneous helicon 
wave is more complicate. It includes the oscillating of the electrons relative to 
ions and the rotating of the electron component relative to ions with radially inhomogeneous 
stationary angular velocity $\Omega\left(R_{e0}\right)$. Both these motions are included in Eq. 
(\ref{42}), solution of which may be derived only numerically. 

\section{The ion acoustic instability of the cylindrical plasma in the field of the azimuthally 
symmetric helicon wave}\label{sec4} 

For understanding the qualitative and the quantitative effect of the rotation of electrons 
with angular velocity $\Omega_{e}\left(R_{e0}\right)\ll \omega_{0}$ on the plasma stability we 
derive the simplest analytical solution to Eq. (\ref{42}) in which terms with $p\neq p_{1}\neq 
0$ are neglected. Equation (\ref{42}) with $n=p=p_{1}=0$ in this case becomes 
\begin{eqnarray}
&\displaystyle
\Phi_{m}\left(k_{\bot},k_{z},\omega\right)\varepsilon_{m}\left(k_{\bot},k_{z}, \omega\right)=0,
\label{46}
\end{eqnarray}
where 
\begin{eqnarray}
&\displaystyle
\varepsilon_{m}\left(k_{\bot},k_{z}, \omega\right)=1-\frac{\omega^{2}_{pi}}{\omega^{2}}
+ J^{2}_{0}\left(ma_{e}\left(R_{e0}\right)\right)
\frac{1}{k^{2}\lambda^{2}_{De}}\left\{1
\right.
\nonumber
\\  
&\displaystyle
\left.
+ i\sqrt{\pi}\left(\omega+m\Omega_{e}\left(R_{e0}\right)-m\omega_{de}\left(1-\frac{1}{2}
\eta_{e}\right)\right)\right.
\nonumber
\\  
&\displaystyle
\left.\times\frac{1}{\sqrt{2}k_{z}v_{Te}}W\left(z_{e0}\right)I_{0}\left(k_{\bot}^{2}\rho^{2}_{e}
\right)e^{-k_{\bot}^{2}\rho^{2}_{e}}
\right.
\nonumber
\\  
&\displaystyle
\left.
-\eta_{e}\frac{m\omega_{de}}{\sqrt{2}k_{z}v_{Te}}e^{-k_{\bot}^{2}\rho^{2}_{e}}
\left[z_{e0}\left(1+i\sqrt{\pi}z_{e0}
W\left(z_{e0}\right)\right)I_{0}\left(k_{\bot}^{2}\rho^{2}_{e}\right)\right.
\right.
\nonumber
\\  
&\displaystyle
\left.\left.-i\sqrt{\pi}W\left(z_{e0}\right)k_{\bot}^{2}\rho^{2}_{e}\left(I_{0}\left(k_{\bot}^{2}
\rho^{2}_{e}\right)
-I_{1}\left(k_{\bot}^{2}\rho^{2}_{e}\right)\right)\right]\right\},
\label{47}
\end{eqnarray}
in which
\begin{eqnarray}
&\displaystyle
z_{e0}=\frac{\omega+m\Omega_{e}\left(R_{e0}\right)}{\sqrt{2}k_{z}v_{Te}}.
\label{48}
\end{eqnarray}
The solution to Eq. (\ref{46}) is
\begin{eqnarray}
&\displaystyle
\Phi_{m}\left(k_{\bot},k_{z},\omega\right)= \Phi_{m}\left(k_{\bot},k_{z},\omega_{m}
\left(k_{\bot},k_{z}\right)\right)\delta\left(\omega-\omega_{m}\left(k_{\bot},k_{z}\right)\right) 
\label{49}
\end{eqnarray}
for $ \omega=\omega_{m}\left(k_{\bot},k_{z}\right)$ and $\Phi_{m}\left(k_{\bot},k_{z},\omega\right) 
=0$ for $ \omega\neq\omega_{m}\left(k_{\bot},k_{z}\right)$, 
where $\omega_{m}\left(k_{\bot},k_{z}\right)$ is the solution to the equation 
\begin{eqnarray}
&\displaystyle
\varepsilon_{m}\left(k_{\bot},k_{z}, \omega\right)=0.
\label{50}
\end{eqnarray}
The inverse Fourier-Bessel transform of solution (\ref{49})
\begin{eqnarray}
&\displaystyle
\Phi_{m}\left(r, z, t\right)=\int dk_{z}dk_{\bot}\Phi_{m}\left(k_{\bot},k_{z},\omega\right)\delta
\left(\omega-\omega_{m}\left(k_{\bot},k_{z}\right)\right)
J_{m}\left(k_{\bot}r\right)e^{ik_{z}z-i\omega t} 
\label{51}
\end{eqnarray}
for the separate Fourier-Bessel harmonic with wave numbers $\hat{k}_{\bot}$ and $\hat{k}_{z}$, 
\begin{eqnarray}
&\displaystyle
\Phi_{m}\left(k_{\bot},k_{z}, \omega\right)=\Phi_{m}\left(\hat{k}_{\bot},\hat{k}_{z}
\right)\delta\left(k_{\bot}-\hat{k}_{\bot}\right)\delta\left(k_{z}-\hat{k}_{z}\right)\delta
\left(\omega-\omega_{m}\left(\hat{k}_{\bot},\hat{k}_{z}\right)\right),
\label{52}
\end{eqnarray}
of the $\Phi_{m}\left(k_{\bot},k_{z}, \omega\right)$ spectrum gives 
\begin{eqnarray}
&\displaystyle
\Phi_{m}\left(r, z, \omega\right)=\Phi_{m}J_{m}\left(\hat{k}_{\bot}r\right)e^{i\hat{k}_{z}z
-i\omega_{m}\left(\hat{k}_{\bot},\hat{k}_{z}\right) t}.
\label{53}
\end{eqnarray}
Thus, Eq. (\ref{50}), derived under condition (\ref{39}), determines the dispersive 
properties of the short scale  radially inhomogeneous cylindrical waves 
with a radial profile determined by the Bessel function $J_{m}\left(k_{0\bot}r\right)$. 
Equation (\ref{50}) accounts for the coupled effect of the electron diamagnetic drift 
caused by the plasma density and electron temperature inhomogeneity of the radially 
inhomogeneous plasma with a 
cylindrical geometry, and of the electrons rotation with angular velocity 
$\Omega_{e}\left(R_{e0}\right)$. The simple analytical solution to Eq. (\ref{50})
may be derived for the case $\eta_{e}=0$ of a plasma with homogeneous 
electron temperature. For this case Eq. (\ref{50}) becomes
\begin{eqnarray}
&\displaystyle
\varepsilon_{m}\left(k_{\bot},k_{z}, \omega\right)=1-\frac{\omega^{2}_{pi}}{\omega^{2}}
\nonumber
\\  
&\displaystyle
+ J^{2}_{0}\left(ma_{e}\left(R_{e0}\right)\right)
\frac{1}{k^{2}\lambda^{2}_{De}}
\left[ 1+ i\sqrt{\frac{\pi}{2}}\frac{\omega+m\hat{\Omega}_{e}\left(R_{e0}\right)}{k_{z}
v_{Te}}W\left(z_{e}\right)A_{e0}\right]=0.
\label{54}
\end{eqnarray}
where $A_{e0}=I_{0}\left(k_{\bot}^{2}\rho^{2}_{e}\right)e^{-k_{\bot}^{2}\rho^{2}_{e}}$. 
The frequency $\hat{\Omega}_{e}\left(R_{e0}\right)$, 
is determined as
\begin{eqnarray}
&\displaystyle
\hat{\Omega}_{e}\left(R_{e0}\right)=\Omega_{e}\left(R_{e0}\right)-\omega_{de}\left(R_{e0}\right).
\label{55}
\end{eqnarray}
The solution to Eq. (\ref{54}) for the $\left(|z_{e}|\ll 1\right)$ is 
$\omega\left(\mathbf{k}\right)=\omega_{s}+\delta \omega\left(\mathbf{k}\right)$, where $
\omega_{s}\left(\mathbf{k}\right)$ is the frequency of the ion acoustic wave, 
\begin{eqnarray}
&\displaystyle
\omega^{2}_{s}
\left(\mathbf{k}\right)=k^{2}v^{2}_{s}\left(J^{2}_{0}\left(ma_{e}\left(R_{e0}\right)\right)
+k^{2}\lambda^{2}_{De}\right)^{-1},
\label{56}
\end{eqnarray}
$v_{s}=\left(T_{e}/m_{i}\right)^{1/2}$ is the ion acoustic velocity, and $\delta \omega
\left(\mathbf{k}\right)$ 
with an accuracy to terms on the order of $\left(\delta \omega\left(\mathbf{k}\right)/\omega_{s}
\right)^{2}\ll 1$ is 
\begin{eqnarray}
&\displaystyle \delta\omega\left(\mathbf{k}\right)=-\frac{i\sqrt{\pi}}{2}\omega_{s}
\frac{z_{e0}J^{2}_{0}\left(ma_{e}\left(R_{e0}\right)\right)}{\left(J^{2}_{0}\left(ma_{e}\left(R_{e0}
\right)\right)
+k^{2}\lambda^{2}_{De}\right)}W\left(z_{e0}\right)A_{e0}
\label{57}
\end{eqnarray}
with $z_{e0}=\left(\omega_{s}\left(\mathbf{k}\right)+m\hat{\Omega}_{e}
\left(R_{e0}\right)\right)/\sqrt{2}k_{z}v_{Te}$, where $|z_{e0}|< 1$ 
when $k_{z}/k>\sqrt{m_{e}/m_{i}}$. 
The ion acoustic instability develops when $z_{e0}<0$, that occurs when 
\begin{eqnarray}
&\displaystyle
-m\hat{\Omega}_{e}\left(R_{e0}\right)>\omega_{s},
\label{58}
\end{eqnarray}
with the growth rate $\gamma_{s}\left(\mathbf{k}\right)=\text{Im}\,\delta\omega\left(\mathbf{k}
\right)$ equal to 
\begin{eqnarray}
&\displaystyle \gamma_{s}\left(\mathbf{k}\right)=\text{Im}\,\delta\omega\left(\mathbf{k}
\right) \approx -\frac{\sqrt{\pi}}{2}\frac{\omega_{s}\left(\mathbf{k}\right)z_{e0}J^{2}_{0}
\left(ma_{e}\left(R_{e0}\right)\right)}
{\left(J^{2}_{0}\left(ma_{e}\left(R_{e0}\right)\right)+k^{2}
\lambda^{2}_{De}\right)}e^{-z_{e0}^{2}}A_{e0}.
\label{59}
\end{eqnarray}
Note, that because $R_{e0}=m/k_{\bot}$, condition (\ref{58}) may be presented in the form \\ 
$-R_{e0}\hat{\Omega}_{e}\left(R_{e0}\right)>v_{s}\left(J^{2}_{0}\left(ma_{e}\left(R_{e0}\right)\right)
+k^{2}\lambda^{2}_{De}\right)^{-1/2}$, i. e. the "azimuthal electron 
current velocity" should be larger than the ion acoustic velocity. Another presentation of these 
results may be given by the introduction, instead of "generalized" angular velocity 
$\hat{\Omega}_{e}\left(R_{e0}\right)$, the "generalized" drift frequency 
\begin{eqnarray}
&\displaystyle
\omega^{*}_{de}\left(R_{e0}\right)=\omega_{de}\left(R_{e0}\right)-\Omega_{e}\left(R_{e0}\right)=
-\hat{\Omega}_{e}\left(R_{e0}\right),
\label{60}
\end{eqnarray}
which accounts for the total electron drift caused by radial inhomogeneity of 
the electron density and of the helicon wave. 
With drift frequency $\omega^{*}_{de}\left(R_{e0}\right)$, \\ $z_{e0}=
\left(\omega_{s}\left(\mathbf{k}\right)
-m\omega^{*}_{de}\left(R_{e0}\right)\right)/\sqrt{2}k_{z}v_{Te}$, 
and the condition (\ref{58}) for the ion acoustic instability development becomes 
\begin{eqnarray}
&\displaystyle
m\omega^{*}_{de}\left(R_{e0}\right)>\omega_{s}.
\label{61}
\end{eqnarray} 
This condition is the same as for the ion acoustic instability of the radially 
inhomogeneous plasma\cite{Kadomtsev} without the helicon wave.

It is instructive to derive the numerical estimates for the frequency (56) 
and for the growth rate (59) of the considered ion acoustic instability. As a sample, 
we consider stability at $R_{e0}=2.5$cm of the cylindrical argon plasma with density $n_{e0}
\left(R_{e0}\right)= 10^{12}$\,cm$^{-3}$, the electron temperature $T_{e}=4$\,eV, the ion 
temperature 
$T_{i}=2.6\cdot 10^{-2}$\,eV in the magnetic field $B_{0}=50$\,mT and in the electric field 
$E_{1r}\sim E_{1\varphi}\sim 5$\,V/cm of the helicon wave with frequency 
$\omega_{0}=10 ^{7}$\,s$^{-1}$. For these plasma parameters
$\rho_{i}\approx 2$\,cm, $\rho_{e}\approx 0,1$\,cm, $v_{Te}=8.4\cdot 10^{7}$\,cm/s, $\lambda_{De}
=1.5\cdot 10^{-3}$ cm, 
$v_{s}=3\cdot10^{5}$cm$\cdot$ s$^{-1}$, $\omega_{de}=2.2 \cdot 10^{6}$s$^{-1}$  for plasma density 
inhomogeneity length $l_{n}=\left(\partial\ln n_{0e}\left(R_{e0}\right)/R_{e0}\partial R_{e0}
\right)^{-2}=2$\,cm, $\Omega_{e}\left(R_{e0}\right)=8\cdot 10^{5}$\,s$^{-1}$  and 
$\omega^{*}_{de}\left(R_{e0}\right)=3\cdot 10^{6}$\,s$^{-1}$. We consider the perturbations 
with $k_{\bot}\rho_{e}=1$ for which $k_{\bot}=10$\,cm$^{-1}$ and the 
azimuthal mode number $m=k_{\bot}R_{e0}=25$. The argument $ma_{e}\left(R_{e0}\right)$ 
of the Bessel function $J_{0}$ is equal to $10$ and $J_{0}\left(10\right)=-0,246$ 
(whereas $J_{1}\left(10\right)
= 0.0435\ll |J_{0}\left(10\right)|$),  the ion acoustic frequency $\omega_{s}=6\cdot 10^{7}$s$^{-1}>
\omega_{0}$.  Because $m\omega^{*}_{de}\left(R_{e0}\right)=7.5\cdot 10^{7}>\omega_{s}$, 
condition (\ref{61}) reveals that the ion acoustic instability  for the presented data develops. 
The magnitude of the instability growth rate depends on the value of the $k_{z}$ wave number. For 
$k_{z}=1.26$ cm$^{-1}$ ($\lambda_{z}=5$cm), $|z_{e}|=0,25$, we derive the estimate $\gamma \approx  
10^{-1}\omega_{s}=6\cdot 10^{6}$s$^{-1}$.

In our theory we consider the model of the collisionless plasma. In the real experimental conditions,
a plasma in helicon sources is partially ionized. The collisions of electrons with neutrals ionize 
neutral gas and through efficient ionization neutral gas density decreases by more than a factor of 
1/10\cite{Gilland}. For our numerical example, the density of the neutral argon gas may be 
estimated by the value $n_{argon}\lesssim 10^{11}$\,cm$^{-3}$. The electron-argon collision frequency
$\nu_{en}=n_{argon}\sigma v_{Te}$, where $\sigma\backsimeq 5\cdot 10^{-15}$\,cm$^{2}$ is the cross 
section of the electron-neutral scattering, for our case is negligible small: 
$\nu_{en}\approx 4,2\cdot 10^{4}$ \,s$^{-1}$ $\ll \gamma \sim 6\cdot 10^{6}$\,s$^{-1}$, that confirms 
the validity of the employed model of the collisionless plasma.   

The maximum growth rate (\ref{59}) attains for $z_{e0}=-1/\sqrt{2}$, and for 
$k_{\bot}\rho_{e}\gg 1$ it is equal to
\begin{eqnarray}
&\displaystyle \gamma_{max}\left(\mathbf{k}\right)\approx 0.054\frac{\left(\omega_{ce}
\omega_{ci}
\right)^{1/2}J^{2}_{0}\left(ma_{e}\left(R_{e0}\right)\right)}{\left(J^{2}_{0}
\left(ma_{e}\left(R_{e0}\right)\right)+k^{2}\lambda^{2}_{De}
\right)^{3/2}}.
\label{62}
\end{eqnarray}
It follows from Eqs. (\ref{45}), (\ref{54}), (\ref{56}) that by the transformations 
\begin{eqnarray}
&\displaystyle
\frac{m}{R_{e0}}\rightarrow k_{y}, \quad m\omega^{*}_{de}\rightarrow k_{y}v_{de}, 
\quad m\hat{\Omega}\rightarrow k_{y}V_{0\bot}, \,\,{\text{and}} \quad J^{2}_{0}
\left(ma_{e}\left(R_{e0}\right)\right) \rightarrow 1
\label{63}
\end{eqnarray}
Eq. (\ref{54}) and its solutions (\ref{56}), (\ref{59}) become equal to the dispersion 
equation for the slab model of the inhomogeneous plasma with electron current flowing 
perpendicularly to a magnetic field and to its solution for the ion acoustic current driven 
instability, respectively\cite{Lashmore-Davies}. By using this similarity of the considered 
microscale ion acoustic instability in the cylindrical and in the slab plasma geometries, we can 
employ the estimates for the energy density $W_{E}=\left(4\pi\right)^{-1}\int d\mathbf{k} k^{2}
\Phi^{2}\left(\mathbf{k}\right)$ of the electric field
\begin{eqnarray}
&\displaystyle
\frac{W_{E}}{n_{0e}T_{e}}\sim 5\cdot 10^{-4}\frac{\omega_{ce}}{\omega_{pe}}\frac{T_{e}}{T_{i}},
\quad \left(k\lambda_{De}\sim 1, \gamma=\gamma_{max}\right)
\label{64}
\end{eqnarray}
at the saturation state of the instability, resulted from the induced scattering of the ion acoustic 
wave by the unmagnetized ions\cite{Bychenkov}. The interaction of the magnetised electrons with ion 
acoustic turbulence under condition of the Cherenkov resonance results in the growth of the electron 
temperature $T_{e\parallel}$ along the magnetic field determined by the equation  
\begin{eqnarray}
&\displaystyle
n_{e}\frac{dT_{e\parallel}}{dt}\sim \frac{R_{e0}\hat{\Omega}\left(R_{e0}\right)}{v_{s}}
\left(\omega_{ci}\omega_{ce}\right)^{1/2}W_{E}
\sim \nu_{eff}\frac{k^{2}_{0}c^{2}}{\omega^{2}_{pe}}W_{0}\left(R_{e0}\right),
\label{65}
\end{eqnarray}
where $W_{E}$ is determined by Eq. (\ref{64}), and $W_{0}\left(R_{e0}\right)$ is the energy 
density of the helicon wave at radius $R_{e0}$ and $\nu_{eff}$ is the effective collision 
frequency of the electrons with electric field of the ion acoustic turbulence.

\section{Conclusions}\label{sec5}
In this paper, we develop the theory of the microinstabilities of the cylindrical plasma 
excited by the cylindrically symmetric helicon wave with accounting for the cylindrical geometry 
and the radial inhomogeneities of the helicon wave 
and of a plasma. By employing the guiding center coordinates for the cylindrical geometry
we obtain the linear integral equation (\ref{38}) for the  the $m$-th harmonic $\Phi_{m}
\left(k_{\bot},k_{z},\omega\right)$ of the Fourier-Bessel transformed potential of the 
electrostatic perturbations of a helicon source plasma. 
The approximate solution (\ref{40}) 
to Eq. (\ref{38}) for the short scale perturbations
(\ref{39}) is derived. The explicit form (\ref{42}) of this equation is derived for the Maxwellian 
electron distribution. Equation (\ref{42}) is the basic equation for the investigations of the 
parametric and current driven instabilities of the cylindrical plasma in the radially 
inhomogeneous helicon wave. 

The developed theory reveals new macroscale effect of the azimuthal steady rotation of 
electrons with a radially inhomogeneous angular velocity, caused by the radial inhomogeneity of the 
helicon wave. We found, that this effect is responsible for 
the development of the ion acoustic instability driven by the coupled effect of the 
azimuthal steady rotation of electrons and of the electron diamagnetic drift at radius 
$R_{e0}$ where condition (\ref{58}) ( or (\ref{61})) holds. 

\begin{acknowledgments}
This work was supported by National R\&D Program through the National Research Foundation of 
Korea (NRF) funded by the Ministry of Education, Science and Technology (Grant No. 
NRF-2017R1A2B2011106) and BK21 FOUR, the Creative Human Resource Education and Research 
Programs for ICT Convergence in the 4th Industrial Revolution, and by the National 
Research Foundation of Ukraine (Grant No.2020.02/0234).
\end{acknowledgments}

\bigskip
{\bf DATA AVAILABILITY}

\bigskip
The data that support the findings of this study are available from the corresponding author upon 
reasonable request.

\end{document}